\documentclass[conference]{IEEEtran}

\usepackage{amssymb, amsmath}
\usepackage{algorithm}
\usepackage[noend]{algpseudocode}
\usepackage{graphicx}
\usepackage{amsthm,amsmath}
\usepackage{amsfonts}
\usepackage{subcaption}
\usepackage{mathtools}
\usepackage{color}
\usepackage{lipsum}
\newtheorem{defn}{Definition}
\newtheorem{exm}{Example}
\newtheorem{propos}{Proposition}
\newtheorem{prob}{Problem}

\def\bs{\mathcal{S}}

\def\Q{\mathcal{Q}}
\def\T{\mathcal{T}}

\def\R{\mathcal{R}}

\newcommand\blfootnote[1]{%
	\begingroup
	\renewcommand\thefootnote{}\footnote{#1}%
	\addtocounter{footnote}{-1}%
	\endgroup
}
\begin{document}

\title{Detecting Pathogenic Social Media Accounts without Content or Network Structure}

\author{\IEEEauthorblockN{Elham Shaabani, Ruocheng Guo, and Paulo Shakarian}
	Arizona State University\\
	Tempe, AZ\\
	Email: \{shaabani, rguo12, shak\}@asu.edu}

\maketitle

\begin{abstract}
	The spread of harmful mis-information in social media is a pressing problem.  We refer accounts that have the capability of spreading such information to viral proportions as ``Pathogenic Social Media" accounts. These accounts include terrorist supporters accounts, water armies, and fake news writers. We introduce an unsupervised causality-based framework that also leverages label propagation. This approach identifies these users without using network structure, cascade path information, content and user's information. We show our approach obtains higher precision (0.75) in identifying Pathogenic Social Media accounts in comparison with random (precision of 0.11) and existing bot detection (precision of 0.16) methods. \blfootnote{U.S. Provisional Patent 62/628,196. Contact shak@asu.edu for licensing information.}
\end{abstract}

\IEEEpeerreviewmaketitle

\section{Introduction}
The spread of harmful mis-information in social media is a pressing problem.  We refer accounts that have the capability of spreading such information to viral proportions as ``Pathogenic Social Media" (PSM) accounts. These accounts include terrorist supporters accounts, water armies, and fake news writers. These organized groups/accounts spread messages regarding certain topics. They might be multiple people that tweet/retweet through multiple accounts to promote/degrade an idea. This can influence public opinion. Identifying PSM accounts has important applications to countering extremism~\cite{khader2016combating,allcott2017social}, the detection of water armies~\cite{chen2013battling,chen2013best,wang2014detection}  and fake news campaigns~\cite{gupta20131,kang2016fake,allcott2017social}. In Twitter, many of these accounts are social bots.

The PSM accounts that propagate information are key to a malicious information campaign and detecting them is critical to understanding and stopping such campaigns. However, this is difficult in practice. Existing methods rely on message content~\cite{morstatter2016new}, network structure~\cite{cao2012aiding} or a combination of both~\cite{subrahmanian2016darpa,davis2016botornot,dickerson2014using}.  However, reliance on information of this type leads to two challenges.  First, network structure is not always available.  For example, the Facebook API does not make this information available without permission of the users (which is likely a non-starter for PSM accounts).  Second, the use of content often necessitates the training of a new model for a previously unobserved topics. For example, PSM accounts taking part in elections in the U.S. and Europe will likely leverage different types of content. In this paper, we propose a method based on causal analysis to avoid these very problems. The main requirement is an activity log of user's activities and timestamp. Further, as our method does not rely on data used in previous approaches, it is inherently complementary – which allows for future combined methods. 

In this paper, we aim to find PSM users in the \emph{viral cascades}. As viral cascades are so rare, the users that cause them are suspicious accounts. To address these issues, we leverage causal analysis~\cite{suppes1970probabilistic,kleinberg2009temporal}. 
We developed, implemented, and evaluated a framework for identifying PSM accounts. This paper makes the following contributions:

\begin{itemize}
	\item  We proposed a PSM detection framework that does not leverage network structure, cascade path information, content and user's information.
	
	\medspace
	\item We observed that PSM accounts have higher causality values.
	
	\medspace
	\item We introduced a series of causality-based metrics for identifying PSM users - which alone can achieve precision of 0.66.
	
	\medspace
	\item We introduced an unsupervised label propagation framework that, when combined with our causal metrics, provide a precision of 0.75. We showed that our framework significantly outperforms random method~(0.11), the content-based bot detection (0.13), all features (0.16), and Sentimetrix~\cite{subrahmanian2016darpa}~(0.11).
	
	\medspace
	\item Our framework is able to find the more important PSM accounts in comparison with the baseline methods. The larger the cascade is, the more important its PSM accounts are and our model can capture those cascades better.

\end{itemize}

The rest of the paper is organized as follows. In Section~\ref{sec:techapp}, we describe our framework that leverages causal analysis and label propagation. Then we present the algorithms in Section~\ref{sec:algo}. This is followed by a description of our dataset in Section~\ref{sec:isisdata}. In Section~\ref{sec:caus_anal}, the causality analysis is discussed. Then we describe our implementation and discuss our results in Section~\ref{sec:res}. Finally, related work is reviewed in Section~\ref{sec:rw}.

\section{Technical Approach} \label{sec:techapp}

\subsection{Technical Preliminaries}
Throughout this paper we shall represent cascades as an ``action log" ($Actions$) of tuples where each tuple $(u,m,t) \in Actions$ corresponds with a user $u \in U$ posting message $m \in M$ at time $t \in T$, following the convention of~\cite{goyal2010learning}. We assume that set $M$ includes posts/repost of a certain original tweet or message. For a given message, we  only consider the first occurrence of each user. We define $Actions_m$ as a subset of $Actions$ for a specific message $m$. Formally, we define it as $Actions_m = \{(u', m', t') \in Actions \ s.t.\ m'=m\}$.

\begin{defn} \emph{\textbf{($m$-participant).}}\label{def:cascade} 
	For a given $m\in M$, user~$u$ is an \textbf{$m$-participant} if there exists $t$ such that $(u,m,t) \in Actions$.
\end{defn}

Note that the users posting tweet/retweet in the early stage of cascades are the most important ones since they play a significant role in advertising the message and making it viral. For a given $m\in M$, we say $m$-participant $i$ ``precedes" $m$-participant $j$ if there exists $t<t'$ where $(i,m,t), (j,m,t') \in Actions$. Thus, we define \textit{key users} as a set of users adopting a message in the early stage of its life span. We formally define \textit{key user} as follows:
\begin{defn}\emph{\textbf{(Key User).}}\label{def:kusers} 
	For a given message $m$, $m$-participant $i$, and $Actions_m$, we say user $i$ is a \textbf{key user} iff user $i$ precedes at least $\phi$ fraction of $m$-participants (formally: $ |Actions_m| \times \phi \le |\{j |  \exists t': (j, m, t') \in \ Actions_m\, \wedge\, t' > t \}|$, $(i, m, t) \in Actions_m$), where $\phi \in (0,1)$.
\end{defn}
The notation  $|\cdot|$ denotes the cardinality of a set. All messages are not equally important. That is, only a small portion of them gets popular. We define \textit{viral messages} as follows:

\begin{defn} \emph{\textbf{(Viral Messages).}}\label{def:cascade} 
	For a given threshold $\theta$, we say that a message $m \in M$ is \textbf{viral} iff $|Actions_m| \ge \theta$. We use $M_{vir}$ to denote the set of viral messages. 
\end{defn}

The Definition~\ref{def:cascade} allows us to compute the prior probability of a message (cascade) going viral as follows:
\begin{equation}
	\rho = \frac{|M_{vir}|}{|M|}
\end{equation}

We also define the probability of a cascade $m$ going viral given some user $i$ was involved as:

\begin{equation}
	p_{m | i} = \frac{|\{m \in M_{vir}\, s.t.\ i\ is\  a \  key \ user\}|}{|\{m \in M\, s.t.\ i\ is\  a \  key \ user\}|}
\end{equation}

We are also concerned with two other measures. First, the probability that two users $i$ and $j$ tweet or retweet viral post $m$ chronologically, and both are key users. In other words, these two users are making post $m$ viral.
\begin{equation}\label{eq:rule1}
	p_{i,j} = 
	\frac{
		\splitfrac{
			|\{m \in M_{vir} | \exists t,t'\, where\, t<t'\, and
		}
		{
			\, (i,m,t),(j,m,t') \in Actions\}|
		}
	}
	{
		|{m \in M | \exists t,t'\, where\, (i,m,t),(j,m,t') \in Actions}|
	}
\end{equation}

Second, the probability that key user $j$ tweets/retweets viral post $m$ and user $i$ does not tweet/retweet earlier than $j$. In other words, only user $j$ is making post $m$ viral.
\begin{equation}\label{eq:rule2}
	p_{\neg{i},j} = \frac{
		\splitfrac{
			|\{m \in M_{vir} | \exists t'\, s.t.\, (j,m,t') \in Actions\, and
		}
		{
			\, \not\exists t\  where\ t <t',\ (i,m,t) \in Actions\}|
		}
	}
	{
		\splitfrac{
			|\{m \in M | \exists t'\, s.t.\, (j,m,t') \in Actions\, and
		}
		{
			\, \not\exists t\  where\ t<t', \ (i,m,t) \in Actions\}|
		}
	}
\end{equation}

Knowing the action log, we aim to find a set of pathogenic social media (PSM) accounts. These users are associated with the early stages of large information cascades and, once detected, are often deactivated by a social media firm. In the causal framework, we introduce a series of causality-based metrics for identifying PSM users.

\subsection{Causal Framework}
We adopt the causal inference framework previously introduced in~\cite{suppes1970probabilistic,kleinberg2009temporal}. We expand upon that work in two ways: (1.) we adopt it to the problem of identifying PSM accounts and (2.) we extend their single causal metric to a set of metrics. Multiple causality measurements provide a stronger determination of significant causality relationships. 
For a given viral cascade, we seek to identify potential users who likely \textit{cause} the cascade viral. We first require an initial set of criteria for such a causal user.  We do this by instantiating the notion of Prima Facie causes to our particular use case below:

\begin{defn} \emph{\textbf{(Prima Facie Causal User).}}\label{def:pf} 
	A user $u$ is a prima facie causal user of cascade $m$ iff: User $u$ is a key user of $m$,  $m \in M_{vir}$, and $p_{m|u} > \rho$.
\end{defn}

For a given cascade $m$, we will often use the language \textit{prima facie causal user} to describe user $i$ is a prima facie cause for $m$ to be viral. In determining if a given prima facie causal user is causal, we must consider other ``related" users. In this paper, we say $i$ and $j$ are $m$-related if (1.) $i$ and $j$ are both prima facie causal users for $m$, (2.) $i$ and $j$ are both key users for $m$, and (3.) $i$ precedes $j$. Hence, we will define the set of ``related users" for user $i$ (denoted $R(i)$) as follows:
\begin{equation}\label{eq:R}
	R(i) = \{j \ s.t. \ j \not= i\ , \exists m \in M \ s.t. \ i,j \ are \ m-related\}
\end{equation}

Therefore, $p_{i,j}$  in (\ref{eq:rule1}) is the probability that cascade $m$ goes viral given both users $i$ and $j$, and $p_{\neg i,j}$ in (\ref{eq:rule2}) is the probability that cascade $m$ goes viral given key user $j$ tweets/retweets it while key user $i$ does not tweet/retweet $m$ or precedes $j$. The idea is that if $p_{i,j} - p_{\neg i,j} > 0$, then user $i$ is more likely a cause than $j$ for $m$ to become viral. We measure \textit{Kleinberg-Mishra causality} ($\epsilon_{K\&M}$) as the average of this quantity to determine how causal a given user $i$ is as follows:
\begin{equation}
	\epsilon_{K\&M}(i) = \frac{\sum_{j\in R(i)} (p_{i,j} - p_{\neg i, j})}{|R(i)|}
\end{equation}
Intuitively, $\epsilon_{K\&M} $ measures the degree of causality exhibited by user $i$.  Additionally, we find it useful to include a few other measures. We introduce \textit{relative likelihood causality} ($\epsilon_{rel}$) as follows: 
\begin{equation}
	\epsilon_{rel}(i) = \frac{\sum_{j\in R(i)} S(i,j)}{|R(i)|}
\end{equation}
\begin{equation}
	S(i, j)=
	\begin{cases}
		(\frac{p_{i,j}}{p_{\neg i, j} + \alpha})-1 , & p_{i,j}> p_{\neg i, j}\\
		0, & p_{i,j} = p_{\neg i,j'}\\
		1- (\frac{p_{\neg i, j}}{p_{i,j}}), &\text{otherwise}
	\end{cases}
\end{equation}

where $\alpha$ is infinitesimal. Relative likelihood causality metric assesses the relative difference between $p_{i,j}$ and $p_{\neg{i},j}$. This helps us to find new users that may not be prioritized by $\epsilon_{K\&M}$.
We also find that if a user is mostly appearing after those with the high value of $\epsilon_{K\&M}$, then it is likely to be a PSM account. One can consider all possible combinations of events to capture this situation. However, this approach is computationally expensive. Therefore, we define $\Q(j) $ as follows:
\begin{equation}\label{eq:Q}
	\Q(j) = \{i \ s.t. \ \ j \in R(i)\}
\end{equation}

Consider the following example:
\begin{exm}
	Consider two cascades (actions) $\tau_1 = \textsc{\{a, b, c, d, e, f, g, h\}}$ and $\tau_2 = \textsc{\{n, m, c, a, h, v, s, t\}}$ where the capital letters signify users. We aim to relate key users while $\phi=0.5$ (Definition~\ref{def:kusers}). Table~\ref{tab:ex1_precond} shows the related users $R(.)$ for each cascade. Note that the final set $R(.)$ for each user, is the union of all sets from the cascades. Set $Q(.)$ for the users of Table~\ref{tab:ex1_precond} are presented in Table~\ref{tab:ex2_precond}.
	\vspace*{-1mm}
	\begin{table}[ht]
		\centering
		\caption{\textmd{Related users $R(.)$ (\ref{eq:R}}) of cascades $\tau_1 = \textsc{\{a, b, c, d, e, f, g, h\}}$ and $\tau_2 = \textsc{\{n, m, c, a, h, v, s, t\}}$ }
		\begin{tabular}{| p{1cm}| c c c|}
			\hline
			\textbf{User} & \textbf{$R_{\tau_1}$} & \textbf{$R_{\tau_2}$} & \textbf{$R$} \\ \hline \hline
			\textsc{a} &\textsc{\{b, c, d, e, f\}} & \textsc{\{h, v\}} & \textsc{\{b, c, d, e, f, h, v\}}\\ \hline
			\textsc{b} & \textsc{\{c, d, e, f\}} &\{\} & \textsc{\{c, d, e, f\}}\\ \hline
			\textsc{c} & \textsc{\{d, e, f\}} & \textsc{\{a, h, v\}} & \textsc{\{a,d, e, f, h, v\}}\\ \hline
			\textsc{d} & \textsc{\{e, f\}} & \{\} & \textsc{\{e, f\}} \\ \hline 
			\textsc{e} & \textsc{\{f\}} &\{\} & \textsc{\{f\}} \\ \hline
			\textsc{n} & \textsc{\{\}} & \textsc{\{m, c, a, h, v\}} & \textsc{\{a, c, h, m, v\}} \\ \hline
			\textsc{m} & \textsc{\{\}} & \textsc{\{c, a, h, v\}} & \textsc{\{a, c, h, v\}} \\ \hline 
			\textsc{h} & \textsc{\{\}} & \textsc{\{v\}} & \textsc{\{v\}}\\ \hline \hline
		\end{tabular}
		\label{tab:ex1_precond}
	\end{table}
	\vspace*{-4mm}
	\begin{table}[ht]
		\centering
		\caption{\textmd{Set $\Q(.)$ of users Table~\ref{tab:ex1_precond}~in (\ref{eq:Q}) }}
		\begin{tabular}{| p{1cm}| c|}
			\hline
			\textbf{User} & \textbf{Total} \\ \hline \hline
			\textsc{a} &\textsc{\{c, n, m\}} \\ \hline
			\textsc{b} & \textsc{\{a\}} \\ \hline
			\textsc{c} & \textsc{\{a, b, n, m\}} \\ \hline
			\textsc{d} & \textsc{\{a, b, c\}} \\ \hline 
			\textsc{e} & \textsc{\{a, b, c, d\}} \\ \hline
			\textsc{n} & \textsc{\{\}} \\ \hline
			\textsc{m} & \textsc{\{n\}} \\ \hline 
			\textsc{h} & \textsc{\{a, c, n, m\}} \\ \hline \hline		
		\end{tabular}
		\label{tab:ex2_precond}
	\end{table}
	
\end{exm}

Accordingly, we define  \textit{neighborhood-based causality}~($\epsilon_{nb}$) as the average $\epsilon_{K\&M}(i)$ for all $i \in Q(j)$ as follows: 
\begin{equation}
	\epsilon_{nb}(j) = \frac{\sum_{i \in \Q(j)} \epsilon_{K\&M} (i)}{|\Q(j)|}
\end{equation}

The intuition behind this metric is that accounts who are retweeting a message that was tweeted/retweeted by several causal users are potential for PSM accounts. We also define the \textit{weighted neighborhood-based causality}~($\epsilon_{wnb}$) as follows:
\begin{equation} \label{eq:weighted}
	\epsilon_{wnb} (j) = \frac{\sum_{i \in \Q(j)} w_i \times \epsilon_{K\&M} (i) }{\sum_{i \in \Q(j)} w_i}
\end{equation}

The intuition behind the metric $\epsilon_{wnb}$ is that the users in $\Q$ may not have the same impact on user $j$ and thus different weights $w_i$ are assigned to each user $i$ with~$\epsilon_{K\&M}(i)$. 

\subsection{Problem Statements}
Our goal is to find the potential PSM accounts from the cascades. Assigning a score to each user and applying threshold-based algorithm is one way of selecting users. In the previous section, we defined causality metrics where each of them or combination of them can be a strategy for assigning scores. Users with high values for causality metrics are more likely to be PSM accounts - later we demonstrate the relationship between these measurements and the real world by identifying accounts deactivated eventually.

\begin{prob}\emph{\textbf{(Threshold-Based Problem).}} \label{prob:1}
	Given a causality metric  $\epsilon_k$ where $k \in \{K\&M, rel, nb, wnb\}$, parameter $\theta$, set of users $U$, we wish to identify set $\{u \ s.t. \ \forall u \in U, \ \epsilon_k(u) \ge \theta\}$.
\end{prob}

We find that considering a set of cascades as a hypergraph where users of each cascade are connected to each other can better model the PSM accounts. The intuition is that densely connected users with high values for causality are the most potential PSM accounts. In other words, we are interested in selecting a user if (1.) it has a score higher than a specific threshold or (2.) it has a lower score but occurs in the cascades where high score users occur. Therefore, we define the \textit{label propagation} problem as follows:

\begin{prob}\emph{\textbf{(Label Propagation Problem).}} \label{prob:2}
	Given a causality metric $\epsilon_k$ where $k \in \{K\&M, rel, nb, wnb\}$, parameters $\theta$, $\lambda$, set of cascades $\T = \{\tau_1, \tau_1, ..., \tau_n\}$, and set of users $U$, we wish to identify set $\bs: \bs_1, \bs_2, ..., \bs_l, ..., \bs_{|U|}$ where $\bs_l = \{u |\forall \tau \in \T, \forall u \in (\tau \backslash \bs_{l-1}), \epsilon_k(u) \ge (H^l_{\tau} - \lambda) \}$ and $H^l_{\tau} = \{min(\epsilon_k(u))  \ s.t. \ \forall u \in \tau \wedge u \in \bigcup\limits_{l' \in [1,l)} \bs_{l'}\}$.
\end{prob}

\section{Algorithms}\label{sec:algo}
\subsection{Algorithm for Threshold-based Problems}
To calculate causality metrics, we use map-reduce programming model. In this approach, we select users with causality value greater than or equal to a specific threshold. We refer to this approach as the \textit{Threshold-Based Selection Approach}. 

\subsection{Label Propagation Algorithms}
\begin{figure}[ht]
\centering
\includegraphics[width=.6\columnwidth]{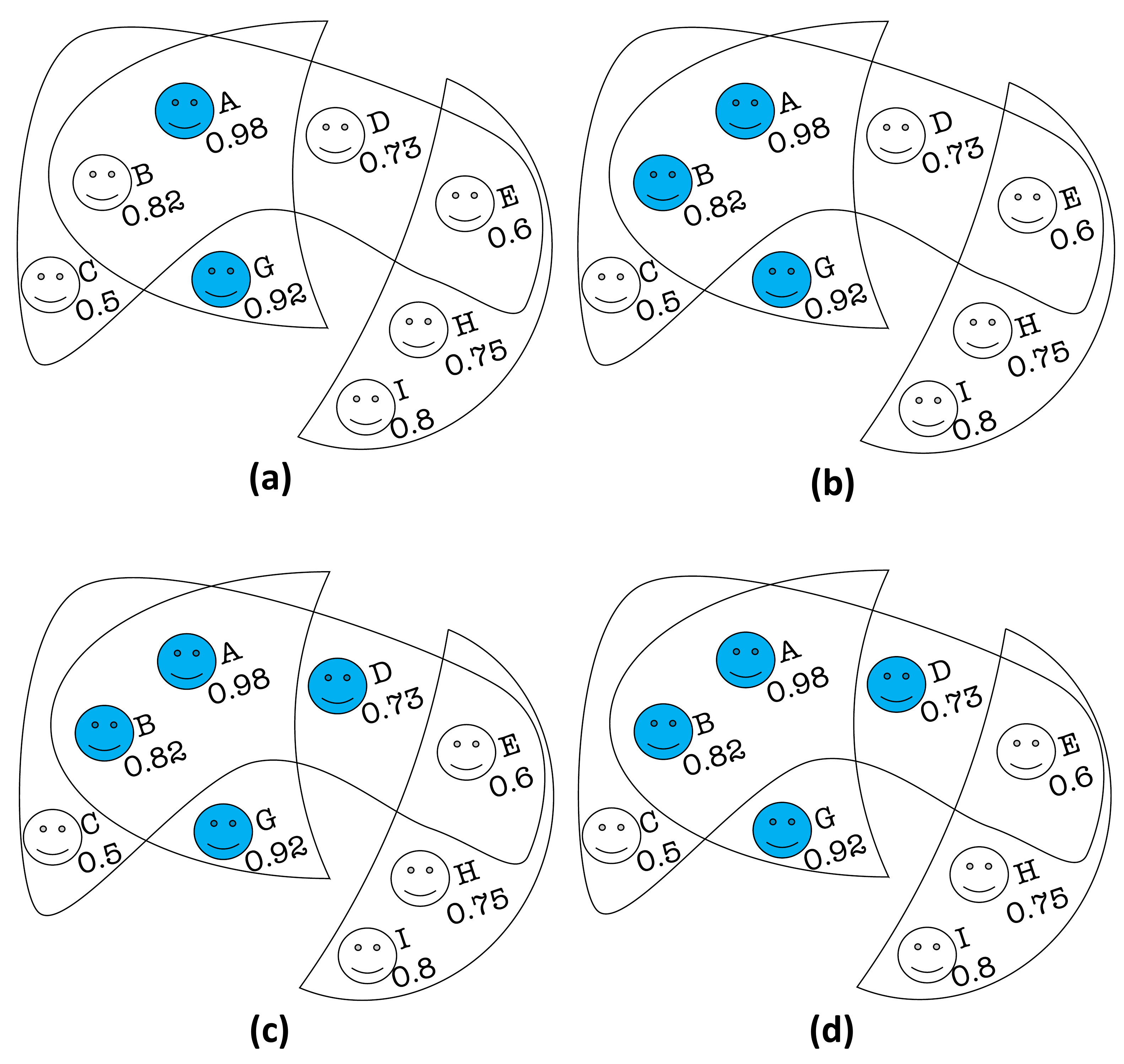}
\caption{A toy example of Algorithm ProSel. Blue faces depict active users.}
\label{fig:exm_alg}
\end{figure}

Label propagation algorithms \cite{zhu2002learning, baluja2008video, raghavan2007near} iteratively propagate labels of a seed set to their neighbors. All nodes or a subset of nodes in the graph are usually used as a seed set. 
We propose a Label Propagation Algorithm (Algorithm~\ref{alg:ProSel}) to solve problem 2. 
We first take users with causality value greater than or equal to a specific threshold (i.e. 0.9) as the seed set. Then, in each iteration, every selected user~$u$ can activate user $u'$ if the following two conditions are satisfied: (1.) $u$ and $u'$ have  at least one cascade (action) in common and (2.) $\epsilon_k(u') \ge \epsilon_k(u) - \lambda, \lambda\in(0,1)$.  Note that, we set a minimum threshold such as 0.7 so that all users are supposed to satisfy it. In this algorithm, inputs are a set of cascades (actions) $\T$, causality metric~$\epsilon_k$ and two parameters $\theta$, $\lambda$ in $(0, 1)$. This algorithm is illustrated by a toy example:

\begin{exm} Consider three cascades \textsc{\{\{a, b, g\}, \{a, b, c, d, e, g, h, i\}, \{e, h, i\}\}} as shown in hypergraph Fig.~\ref{fig:exm_alg}. Let us consider the minimum acceptable value as 0.7; in this case, users \textsc{c} and \textsc{e} would not be activated in this algorithm. Assuming two parameters $\theta = 0.9,\ \lambda = 0.1$, both users \textsc{a} and \textsc{g} get activated (Fig.~\ref{fig:exm_alg}a). Note that an active user is able to activate inactive ones if (1.) it is connected to the inactive user in the hypergraph, (2.) score of the inactive user meets the threshold. In the next step, only user \textsc{b} will be influenced by \textsc{g} ($0.82 \ge 0.92-0.1$) as it is shown in Fig.~\ref{fig:exm_alg}b. Then, user \textsc{d} will be influenced by user \textsc{b} ($0.73 \ge 0.82-0.1$). In the next step (Fig.~\ref{fig:exm_alg}d), the algorithm terminate since no new user is adopted. As it is shown, user \textsc{i} and \textsc{h} are not influenced although they have larger values of $\epsilon$ in comparison with user \textsc{d}.
\end{exm}
\vspace*{-5mm}

\vspace*{-3mm}
\begin{algorithm}
	\caption{\small \textbf{Label Propagaton Algorithm (\emph{ProSel})}}\label{alg:ProSel}
	{
		\begin{algorithmic}[1]
			\Procedure{ProSel}{$\T, \epsilon_k, \theta, \lambda$}
			\State $\bs = \{(u, \epsilon_k(u)) |  \forall u \in U, \epsilon_k(u) \ge \theta\}$
			\State $\R = \bs$
			\State $H = \emptyset$
			\While {$|\bs| > 0$}

			\State $H' = \{(\tau, \epsilon_m) | \forall (\tau, \epsilon) \in H, \ \epsilon_m = min(\epsilon, min(\{\epsilon'= \bs_u \ s.t. \ \forall u \in \tau \wedge u \in \bs\} ))\}$
			\State $H= H' \cup \{(\tau, \epsilon_m) | \forall \tau \in \T \wedge \tau \not \in H', \ \epsilon_m = min(\{\epsilon= \bs_u \ s.t. \ \forall u \in \tau \wedge u \in \bs \} )\}$
			\State $\bs = \{ (u, \epsilon) |\forall \tau \in \T, \ \forall u \in \tau,\ u \not \in \R, \  \epsilon_k(u) \ge (H_\tau - \lambda) \}$
			\State $\R = \R \cup \bs$
			\EndWhile
			\State \textbf{return} $\R$
			\EndProcedure
	\end{algorithmic}}
\end{algorithm}

\begin{propos} Given a set of cascades $\T$, a threshold $\theta$, parameter $\lambda$, and causality values $\epsilon_k$ where $k \in \{K\&M, rel, nb, wnb\}$, ProSel returns a set of users $ \R = \{u | \epsilon_k(u) \ge \theta\ or\ \exists u'\ s.t.\ u',u \in \tau, \epsilon_k(u) \ge \epsilon_k(u') - \lambda\ and\ u'\ is\ picked\}$. Set $\R$ is equivalent to the set $\bs$ in Problem~\ref{prob:2}.
\end{propos}

\begin{propos}
	The time complexity of Algorithm ProSel is $O(|\T| \times avg(log(|\tau|)) \times |U|)$.
\end{propos}

\section{ISIS  Dataset}\label{sec:isisdata}
Our dataset consists of ISIS related tweets/retweets in Arabic gathered from Feb. 2016 to May 2016.  The dataset includes tweets and the associated information such as user ID, re-tweet ID, hashtags, content, date and time. About 53M tweets are collected based on the 290 hashtags such as Terrorism, State of the Islamic-Caliphate, Rebels, Burqa State, and Bashar-Assad, Ahrar Al-Sham, and Syrian Army.  In this paper, we only use tweets (more than 9M) associated with viral cascades. The statistics of the dataset are presented in Table~\ref{tab:dataset} discussed in details below.

\begin{table}[ht]
	\centering
	\caption{\small \textmd{Statistics of the dataset}}
	{
		\begin{tabular}{| p{3cm}| c|}
			\hline
			\textbf{Name} & \textbf{Values} \\ \hline \hline
			Tweets & 9,092,978  \\ \hline 
			Cascades & 35,251  \\ \hline
			Users & 1,249,293 \\ \hline
			Generator users & 8,056 \\ \hline \hline
	\end{tabular}}
	\label{tab:dataset}
\end{table}
\begin{figure}
	\centering
	\begin{subfigure}[b]{.5\columnwidth}
		\includegraphics[width=.95\columnwidth]{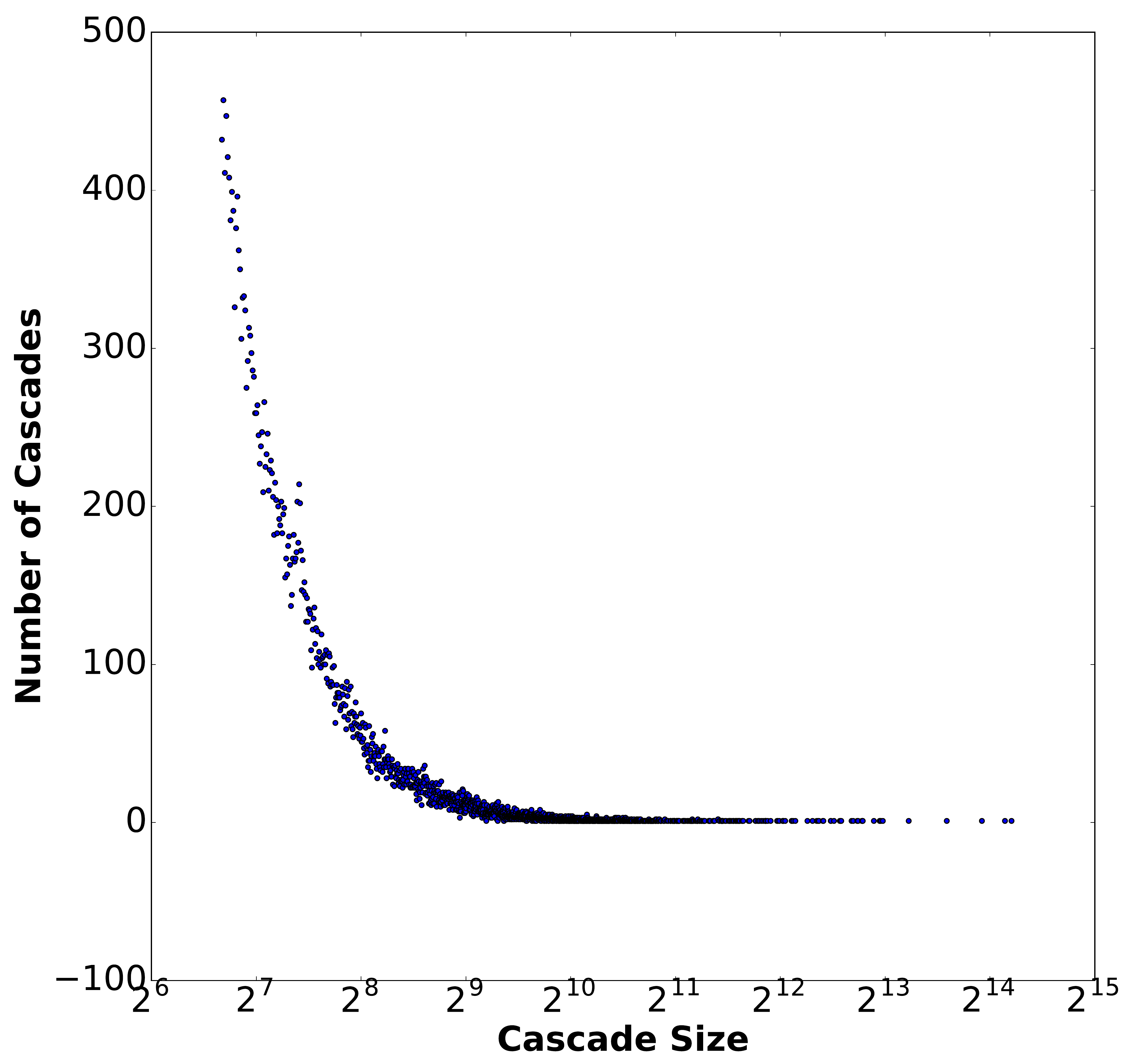}
		\caption{All users} 
		\label{fig:dist_casc_all}
	\end{subfigure}%
	\begin{subfigure}[b]{.5\columnwidth}
		\includegraphics[width=.95\columnwidth]{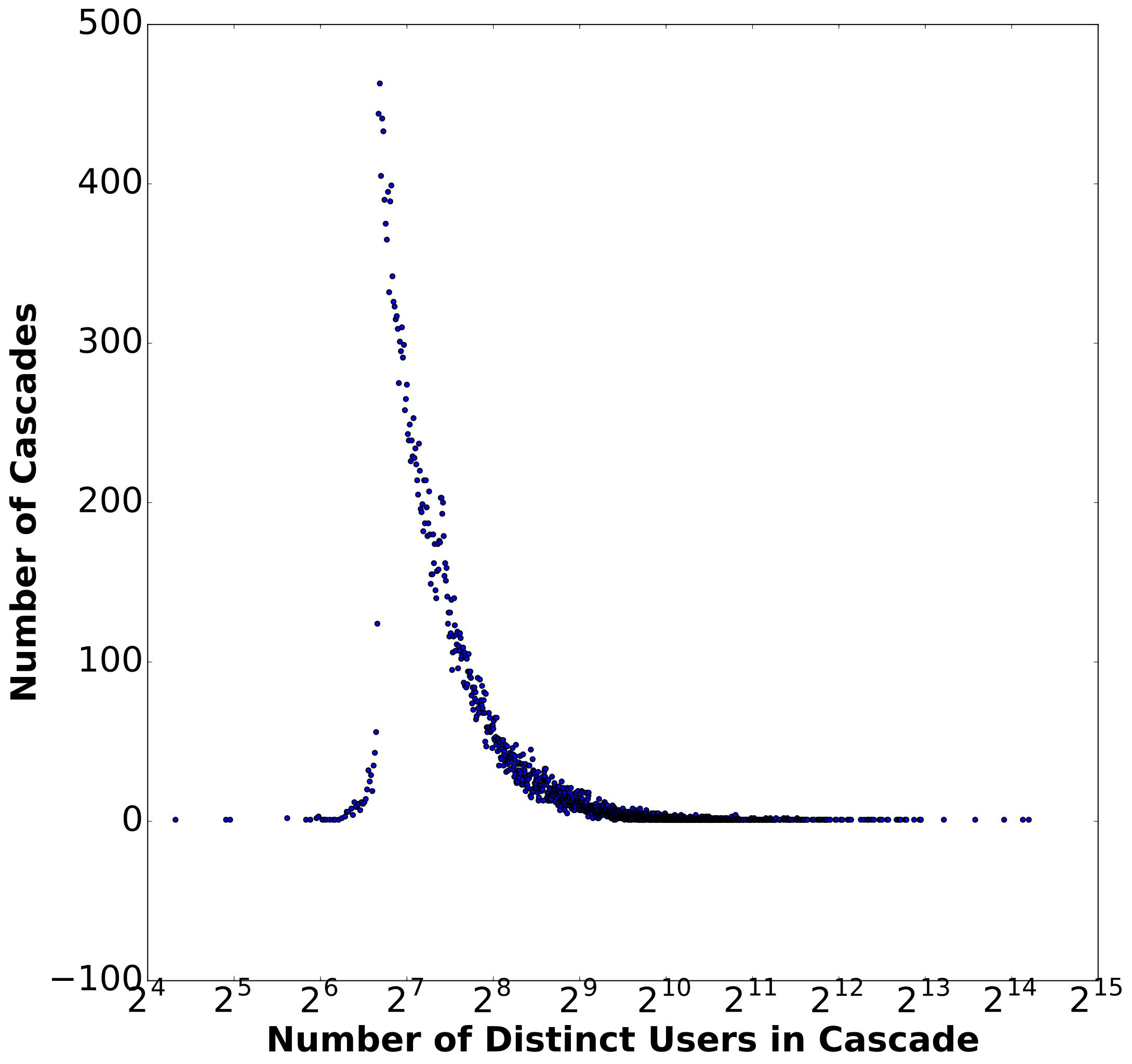}
		\caption{Distinct users}
		\label{fig:dist_casc_uni}
	\end{subfigure}%
	\caption{Distribution of cascades vs cascade size} 
	\label{fig:dist_casc}	
\end{figure}

\begin{figure*}[ht]
	\centering
	\begin{subfigure}[b]{.62\columnwidth}
		\includegraphics[width=1\columnwidth]{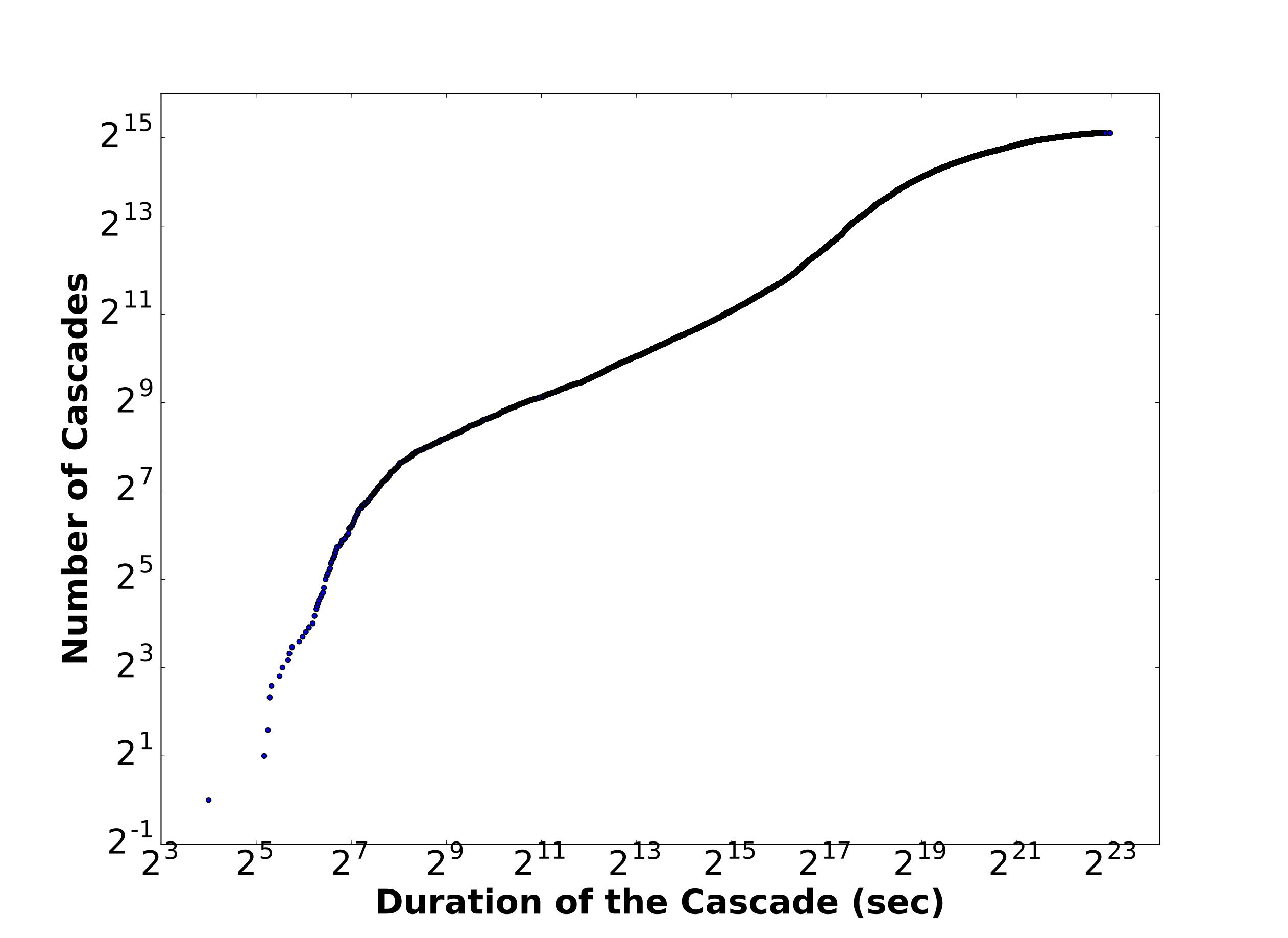}
		\caption{} 
		\label{fig:numCas_DurCas}
	\end{subfigure}%
	\begin{subfigure}[b]{.6\columnwidth}
		\includegraphics[width=1\columnwidth]{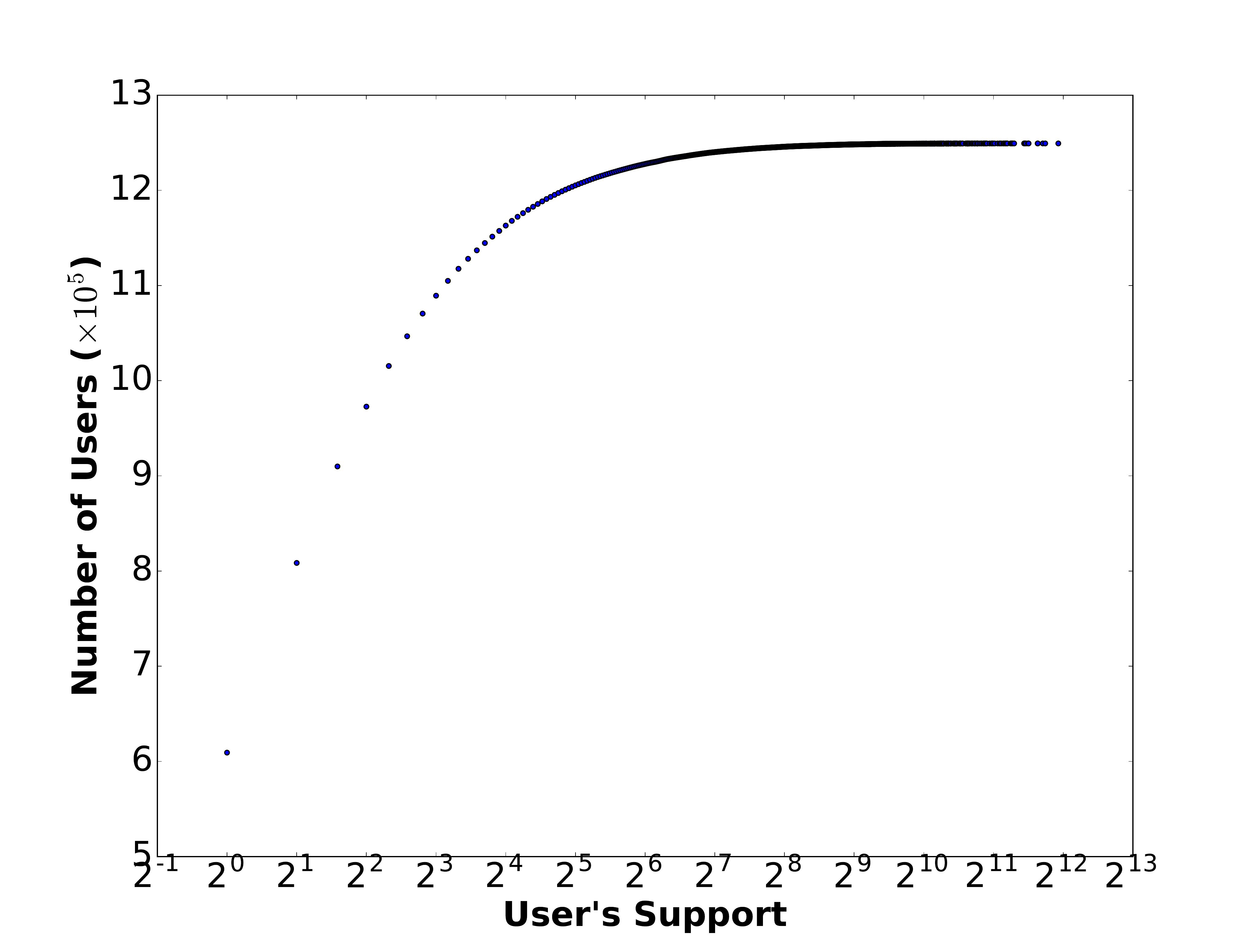}
		\caption{}
		\label{fig:numUs_UsSupp}
	\end{subfigure}%
	\begin{subfigure}[b]{.67\columnwidth}
	\includegraphics[width=1\columnwidth]{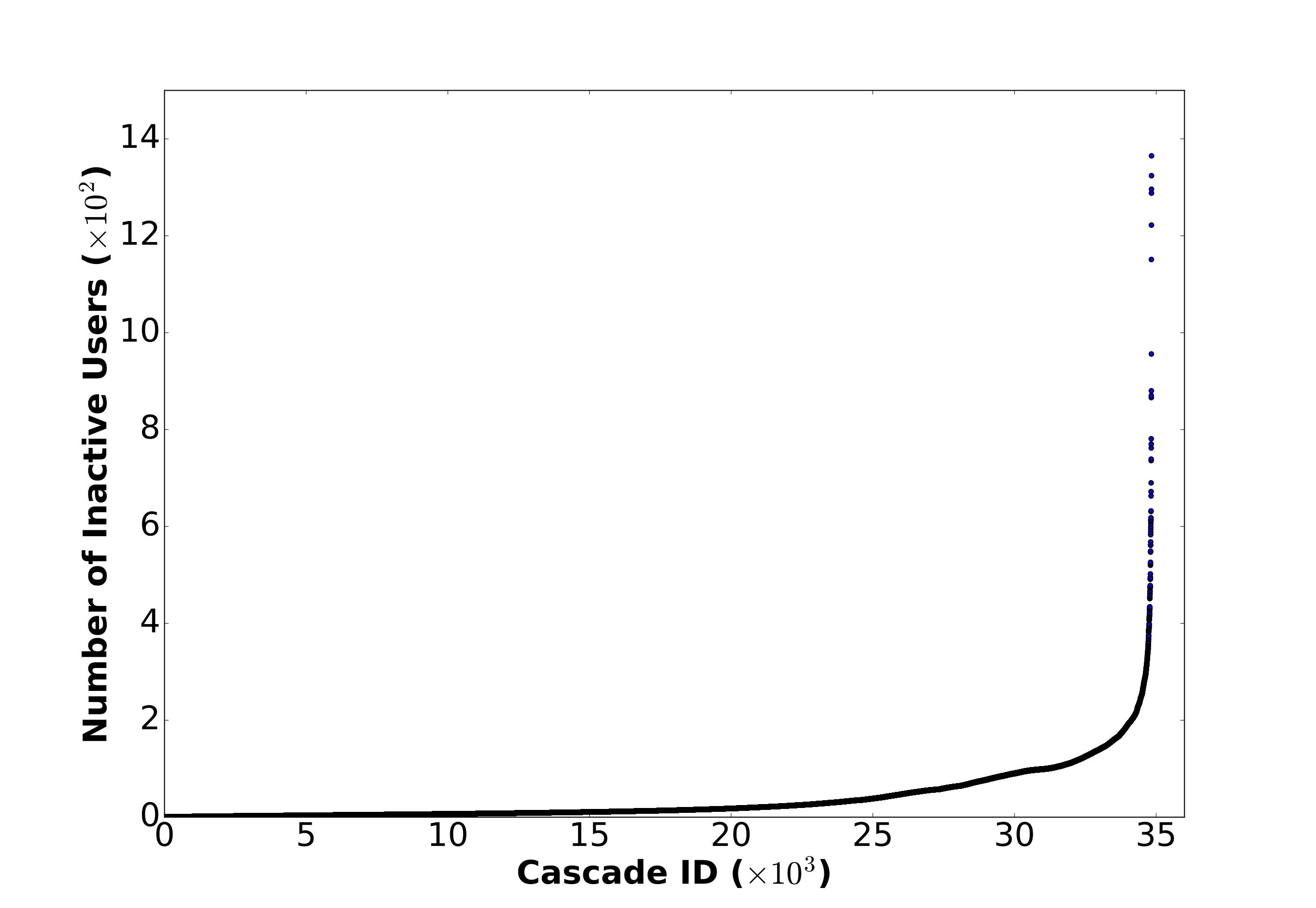}
	\caption{}
	\label{fig:tot_inact_cas}
\end{subfigure}%
	\caption{a) Cumulative distribution of duration of cascades. b) Cumulative distribution of user's occurrence in the dataset. c) Total inactive users in every cascade} 
	\label{fig:dist_casc}	
\end{figure*}

\noindent{\textbf{Cascades.}} In this paper, we aim to identify PSM accounts - which in this dataset are mainly social bots or terrorism-supporting accounts that participate in viral cascades. The tweets that have been retweeted from 102 to 18,892 times. This leads to more than 35k cascades which are tweeted or retweeted by more than 1M users. The distribution of the number of cascades vs cascade size is illustrated in Fig.~\ref{fig:dist_casc_all}. 
There are users that retweet their own tweet or retweet a post several times, we only consider the first tweet/retweet of each user for a given cascade. In other words, duplicate users are removed from the cascades, which make the size of the viral cascades from 20 to 18,789 as shown in Fig.~\ref{fig:dist_casc_uni}. The distribution of the cascades over the cascade life span is illustrated in Fig.~\ref{fig:numCas_DurCas}. Cascades took from 16 seconds to more than 94 days to complete.

\noindent{\textbf{Users.}} There are more than 1M users that have participated in the viral cascades. Fig.~\ref{fig:numUs_UsSupp} demonstrates the cumulative distribution of the number of times a user have participated in the cascades. As it is shown, the larger the support value is, the less number of users exists. Moreover, users have tweeted or retweeted posts from 1 to 3,904 times and on average each user has participated more than 7 times.

\noindent{\textbf{User's Current Status.}} We select \textit{key users} that have tweeted or retweeted a post in its early life span - among first half of the users (according to Definition~\ref{def:kusers}, $\phi = 0.5$), and check whether they are active or not. Accounts are not active if they are suspended or deleted. More than 88\% of the users are active as shown in Table~\ref{tab:active_inactive}. The statistics of the generator users are also reported. Generator users are those that have initiated a viral cascade. As shown, 90\% of the generator users are active as well. Moreover, there are a significant number of cascades with hundreds of inactive users. The number of inactive users in every cascade is illustrated in Fig.~\ref{fig:tot_inact_cas}. Inactive users are representative of automatic and terrorism accounts aiming to disseminate their propaganda and manipulate the statistics of the hashtags of their interest. 

\begin{table}
	\centering
	\caption{\small \textmd{Status of a subset of the users in the dataset.}}
	{
		\begin{tabular}{| p{2.2cm}| c c c|}
			\hline
			\textbf{Name} & \textbf{Active} & \textbf{Inactive} & \textbf{Total}\\ \hline \hline
			Users & 723,727 & 93,770 & 817,497  \\ \hline
			Generator users & 7,243 & 813 & 8,056 \\ \hline \hline
	\end{tabular}}
	\label{tab:active_inactive}
\end{table}

\noindent\textbf{Generator Users.} In this part, we only consider users that have generated (started) the viral tweets.  According to Table~\ref{tab:active_inactive}, there are more than 7k active and 800 inactive generator users. That is, more than 10\% of the generator users are suspended or deleted, which means they are potentially automated accounts. The distribution of the number of tweets generated by generator users shows that most of the users (no matter active and inactive) have generated a few posts (less than or equal to 3) while only a limited number of users are with a large number of tweets. 

\section{Causality Analysis}\label{sec:caus_anal}
Here we examine the behavior of the causality metrics. We analyze users considering their current account status in Twitter. We label a user as active (inactive) if the account is still active (suspended or deleted).

\begin{figure}[h]
	\centering
	\begin{subfigure}[b]{.45\linewidth}	\centering
		\includegraphics[width=.8\linewidth]{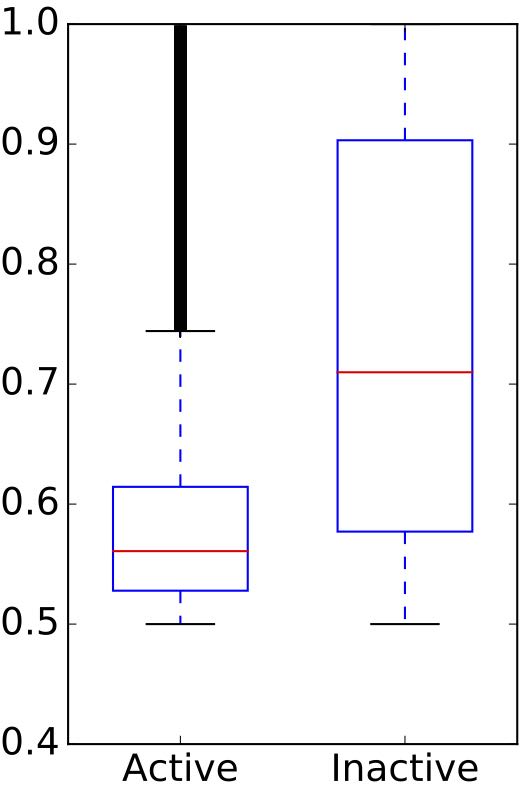}
		\caption{$\epsilon_{K\&M} \ge 0.5$}
		\label{fig:dist_knm_vio}
	\end{subfigure}%
	\begin{subfigure}[b]{0.45\linewidth}	\centering
		\includegraphics[width=.8\linewidth]{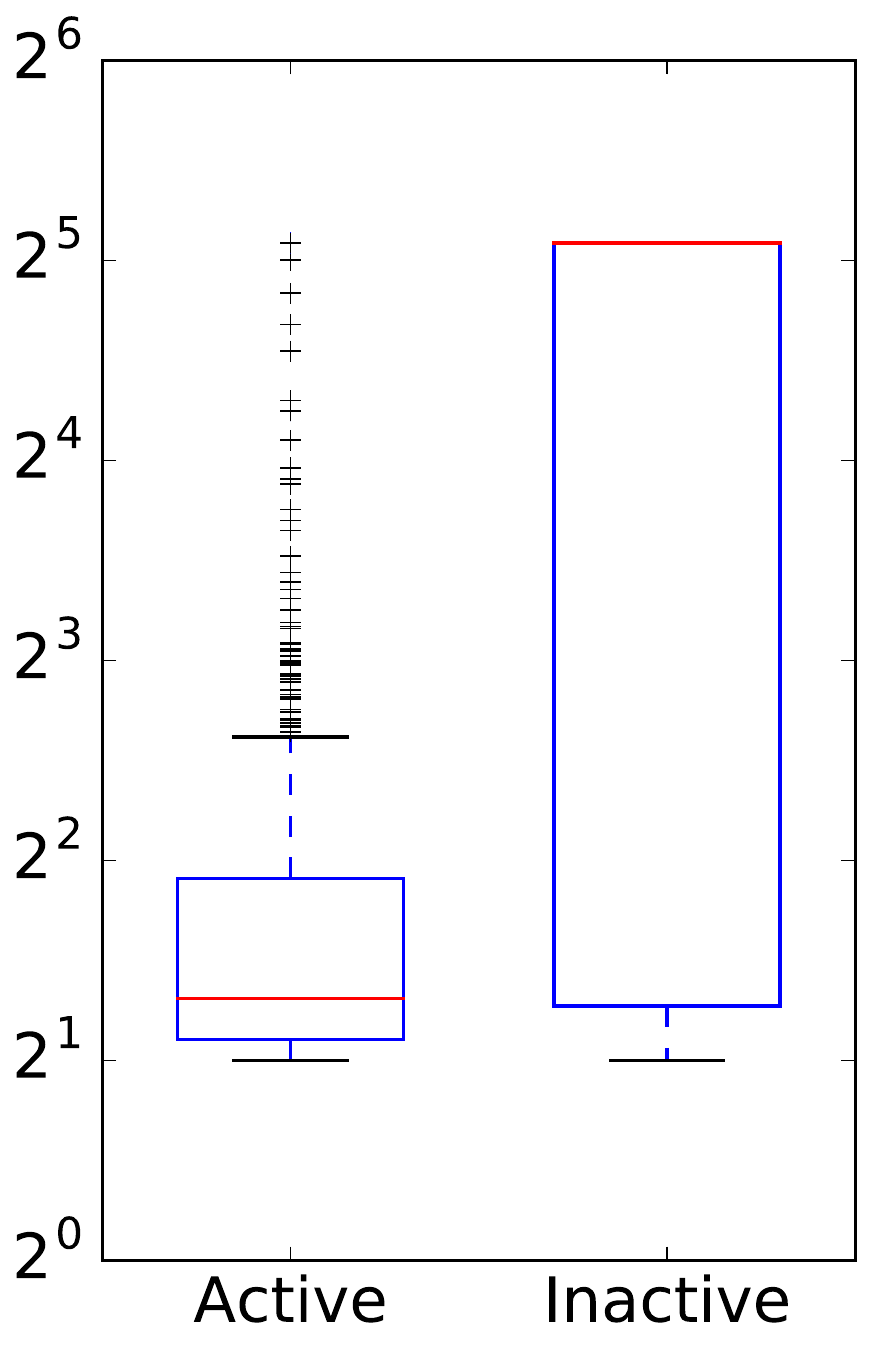}
		\caption{$\epsilon_{rel} \ge 2$}
		\label{fig:dist_rt_vio}
	\end{subfigure}
	\begin{subfigure}[b]{0.45\linewidth}	\centering
		\includegraphics[width=.8\linewidth]{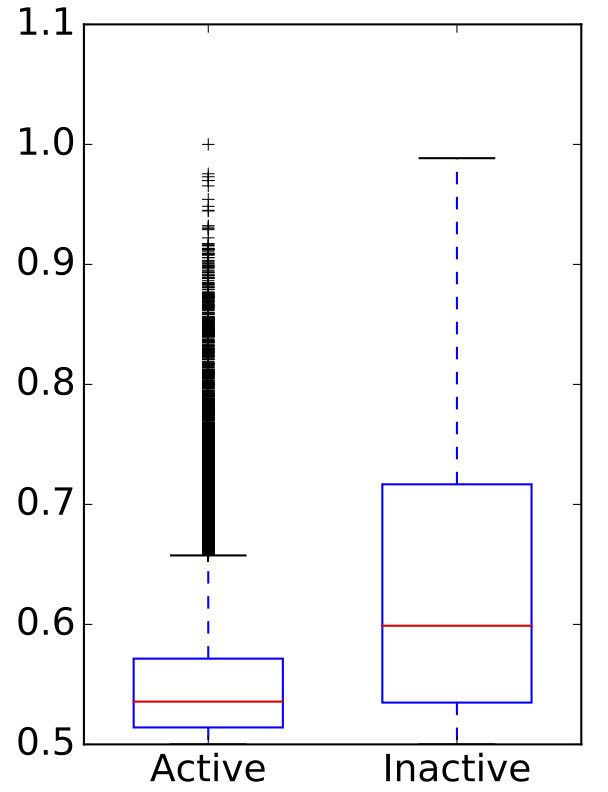}
		\caption{$\epsilon_{nb} \ge 0.5$}
		\label{fig:dist_ngr_vio}
	\end{subfigure}%
	\begin{subfigure}[b]{0.45\linewidth}	\centering
		\includegraphics[width=.8\linewidth]{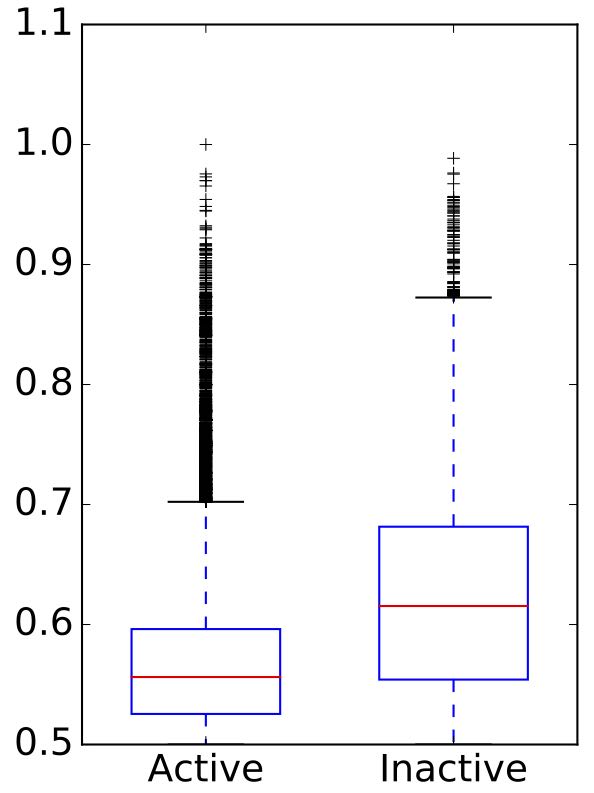}
		\caption{$\epsilon_{wnb} \ge 0.5$}
		\label{fig:dist_ngrw_vio}
	\end{subfigure}
	\caption{Distribution of various causality metrics for active and inactive users.}
\end{figure}

\noindent{\textbf{Kleinberg-Mishra Causality.}} We study the users that get their causality value of $\epsilon_{K\&M}$ greater than or equal to 0.5. As expected, inactive users exhibit different distribution from active users (Fig.~\ref{fig:dist_knm_vio}). We note that significant differences are present - more than 75\% of the active users are distributed between 0.5 and 0.62, while more than 50\% of the inactive users are distributed from 0.75 to 1. Also, inactive users have larger values of mean and median than active ones. Note that number of active and inactive users are 404,536 and 52,452. This confirms that this metric is a good indicator to discriminate PSM users from the normal users. 

\noindent{\textbf{Relative Likelihood Causality.}} This metric magnifies the interval between every pairs of the probabilities that measures the causality of the users; therefore, the values vary in a wide range. Fig.~\ref{fig:dist_rt_vio} displays the distribution of users having relative likelihood causality of greater than or equal to two. In this metric, 1,274 users get very large values. For the sake of readability, very large values are replaced with 34.0 in Fig.~\ref{fig:dist_rt_vio}. More than 50\% of the inactive users get values greater than 32, while the median of active users is 2.48. More than 75\% of the active users are distributed in the range of $(2,4)$. Note that number of active and inactive users in this figure are 3,563 and 1,041, respectively. That is, using this metric and filtering users with the relative likelihood greater than a threshold, leads to the good precision. For example,  the threshold in Fig.~\ref{fig:dist_rt_vio} is set to 2 - the precision is more than 0.22 for inactive class. Considering users with a very large value leads to the precision of more than 0.5 and uncovering a significant number of PSMs - 638 inactive users.

\noindent{\textbf{Neighborhood-Based Causality.}} We study the users that get their causality value of $\epsilon_{nb}$ greater than or equal to 0.5. As expected, inactive users exhibit different distribution from active users as shown in Fig.~\ref{fig:dist_ngr_vio}. Also, inactive users are mostly distributed in the higher range and have larger values of mean and median than active ones. More than 75\% of the active users are distributed between 0.5 and 0.6, while more than 50\% of the inactive users are distributed from 0.6 to 1. Therefore, increasing the threshold results in the higher precision for the PSM users. Note that the number of active and inactive users are 85,864 and 10,165. 

\noindent{\textbf{Weighted Neighborhood-Based Causality.}} This metric is the weighted version of the previous metric ($\epsilon_{nb}$). We assign weight to each user in proportion to her participation rate in the viral cascades. Fig.~\ref{fig:dist_ngrw_vio} shows the distribution of users with $\epsilon_{wnb}$ greater than or equal to 0.5. This metric also displays different distribution for active and inactive users. More than 75\% of the active users are distributed between 0.5 and 0.6, while more than 50\% of the inactive users are distributed from 0.6 to 1. Note that the number of active and inactive users of $\epsilon_{wnb}$ are 52,346 and 16,412. In other words, this metric achieves the largest precision compared to other metrics, 0.24. Clearly, increasing the threshold results in the higher precision for the PSMs.

\section{Results and Discussion}\label{sec:res}
We implement our code in Scala Spark and Python 2.7x and run it on a machine equipped with an Intel Xeon CPU (1.6 GHz) with 128 GB of RAM running Windows 7. We set the parameter $\phi$ to label key users 0.5 (Definition~\ref{def:kusers}). Thus, we are looking for the users that participate in the action before the number of participants gets twice. 

In the following sections, first we look at the existing methods. Then we look at two proposed approaches (see Section~\ref{sec:algo}): (1) \textit{Threshold-Based Selection Approach} - selecting users based on a specific threshold, (2) \textit{Label Propagation Selection Approach} - selecting by applying Algorithm~\ref{alg:ProSel}. The intuition behind this approach is to select a user if it has a score higher than a threshold or has a lower score but occurs in the cascades that high score users exist. We evaluate methods based on true positive (True Pos), false positive (False Pos), precision, the average (Avg CS) and median (Med CS) of cascade size of the detected PSM accounts. Note that in our problem, precision is the most important metric. The main reason is labeling an account as PSM means it should be deleted. However, removing a real user is costly. Therefore, it is important to have a high precison to prevent removing real user.

\subsection{Existing Method}
Here we use the approach proposed by the top-ranked team in the DARPA Twitter Bot Challenge~\cite{subrahmanian2016darpa}. We consider all features that we could extract from our dataset. Our features include tweet syntax (average number of hashtags, average number of user mentions, average number of links, average number of special characters), tweet semantics (LDA topics), and user behaviour (tweet spread, tweet frequency, tweet repeats). We apply three existing methods to detect PSM accounts: 1) \textit{Random} selection: This method achieves the precision of 0.11. This also presents that our data is imbalanced and less than 12\% of the users are PSM accounts. 2) \textit{Sentimetrix (Sentimet.)}: We cluster our data by DBSCAN algorithm. We then propagate the labels from 40 initial users to the users in each cluster based on the similarity metric. We use Support Vector Machines (SVM) to classify the remaining PSM accounts~\cite{subrahmanian2016darpa}. 3) \textit{Classification} methods: In this experiment, we use the same labeled accounts as the previous experiment and apply different machine learning algorithms to predict the label of other samples. We group features based on the limitations of access to data into three categories. First, we consider only using content information (\textit{Content}) to detect the PSM accounts. Second, we use content independent features (\textit{NoCont.})~\cite{subrahmanian2016darpa} to classify users. Third, we apply all features (\textit{All feat.}) to discriminate PSM accounts.  The best result for each setting is when we apply Random Forest using all features. According to the results, this method achieves the highest precision of 0.16. Note that, most of the features used in the previous work and our baseline take advantage of both content and network structure. However, there are situations that the network information and content do not exist. In this situation, the best baseline has the precision of 0.15. We study the average (Avg CS) and median (Med CS) of the size of the cascades in which the selected PSM accounts have participated. Table~\ref{tab:bl} also illustrates the false positive, true positive and precision of different methods.

\begin{table}[ht]
	\centering
	\caption{\small  \textmd{Existing Methods - number of selected users as PSM}}
	{
		\begin{tabular}{|p{1.1cm}| c c c c c|}
			\hline
			\textbf{Method} & \textbf{False Pos} & \textbf{True Pos} & \textbf{Precision} &\textbf{Avg CS}&\textbf{Med CS} \\ \hline \hline
			$Random$ & 80,700 & 10,346 & 0.11 &\textbf{289.99} & \textbf{184}\\ \hline \hline
			$Sentimet.$ &  640,552 & 77,984 & 0.11 &261.37 & 171\\ \hline \hline
			$Content$ &292,039&43,483&0.13&267.66& 174\\ \hline 
			$NoCont.$ &357,027&63,025&0.15& 262.97 &172\\ \hline
			$All feat.$ &164,012&31,131&\textbf{0.16}& 273.21 & 176\\ \hline \hline
			
	\end{tabular}}
	\label{tab:bl}
\end{table}

\subsection{Threshold-Based Selection Approach}\label{sec:res_ths}
In this experiment, we select all the users that satisfy the thresholds and check whether they are active or not. A user is \textit{inactive}, if the account is suspended or closed. Since the dataset is not labeled, we label inactive users as PSM accounts. We set the threshold for all metrics to 0.7 except for relative likelihood causality ($\epsilon_{rel}$), which is set to 7. We conduct two types of experiments: first, we study user selection for a given causality metric. We further study this approach using the combinations of metrics.

\medskip
\noindent{\textbf{Single Metric Selection.}} In this experiment, we attempt to select users based on each individual metric. As expected, these metrics can help us filter a significant amount of active users and overcome the data imbalance issue. Metric $\epsilon_{K\&M}$ achieves the largest recall in comparison with other metrics. However, it has the largest number of false positives. Table~\ref{tab:thr} shows the performance of each metric. The precision value varies from 0.43 to 0.66 and metric $\epsilon_{wnb}$ achieves the best value. Metric $\epsilon_{rel}$ finds the more important PSM accounts with average cascade size of 567.78 and median of 211.  In general, our detected PSM accounts have participated in the larger cascades in comparison with baseline methods. 

We also observe that these metrics cover different regions of the search area. In other words, they select different user sets with little overlap between each other. The common users between any two pairs of the features are illustrated in Table~\ref{tab:thr_common}. Considering the union of all metrics, 36,983 and 30,353 active and inactive users are selected, respectively. 

\begin{table}[ht]
	\centering
	\caption{\small  \textmd{Threshold-based selection approach - number of selected users using single metric}}
	{
		\begin{tabular}{|p{1.1cm}| c c c c c |}
			\hline
			\textbf{Method} & \textbf{False Pos} & \textbf{True Pos} & \textbf{Precision} &\textbf{Avg CS}&\textbf{Med CS}\\ \hline \hline
			$All feat.$ &164,012&31,131&\textbf{0.16} & 273.21 & 176\\ \hline 
			$NoCont.$ &357,027&63,025&0.15& 262.97 &172\\ \hline	\hline
			$\epsilon_{K\&M}$ & 36,159 &  27,192 & 0.43 &383.99 & 178 
			 \\ \hline 
			$\epsilon_{rel}$ & 693 & 641  & 0.48 & \textbf{567.78} & \textbf{211}
			\\ \hline 
			$\epsilon_{nb}$ & 2,268 & 2,927 &  0.56& 369.46 & 183.5 
			\\ \hline 
			$\epsilon_{wnb}$ & 7,463  & 14,409 &\textbf{0.66}& 311.84& 164
			\\ \hline \hline
		\end{tabular}
	}
	\label{tab:thr}
\end{table}
\vspace*{-3mm}
\begin{table}[ht]
	\centering
	\caption{\small  \textmd{Threshold-based selection approach - number of common selected users using single metric}}
	{
		\begin{tabular}{| p{1cm}| c c c | c c c|}
			\hline
			\textbf{Status} & \multicolumn{3}{c|}{\textbf{Active}} &  \multicolumn{3}{c|}{\textbf{Inactive}} \\ \hline
			\textbf{Method} & \textbf{$\epsilon_{rel}$} & \textbf{$\epsilon_{nb}$}  & \textbf{$\epsilon_{wnb}$}  &  \textbf{$\epsilon_{rel}$} & \textbf{$\epsilon_{nb}$}  & \textbf{$\epsilon_{wnb}$} \\ \hline \hline
			$\epsilon_{K\&M}$ & 404 & 1,903 & 6,992  &338 & 2,340 & 11,748 \\ \hline
			$\epsilon_{rel}$  &   & 231  & 175 & & 248 & 229 \\ \hline
			$\epsilon_{wnb}$  &  &  & 1,358 & & & 1,911 \\ \hline  \hline
		\end{tabular}
	}  
	\label{tab:thr_common}
\end{table}

\medskip
\noindent{\textbf{Combination of Metrics Selection.}} According to Table~\ref{tab:thr_common}, most of the metric pairs have more inactive users in common than active users. In this experiment, we discuss if using the combination of these metrics can help improve the performance. We attempt to select users that satisfy the threshold for at least three metrics. We get 1,636 inactive users out of 2,887 selected ones, which works better than $\epsilon_{K\&M}$ and $\epsilon_{rel}$ while worse than $\epsilon_{nb}$ and $\epsilon_{wnb}$. In brief, this approach achieves precision of 0.57. Moreover, the number of false positives (1,251) is lower than most of the other metrics.

\subsection{Label Propagation Selection Approach}
\begin{figure}
	\centering
	\includegraphics[width=.7\columnwidth]{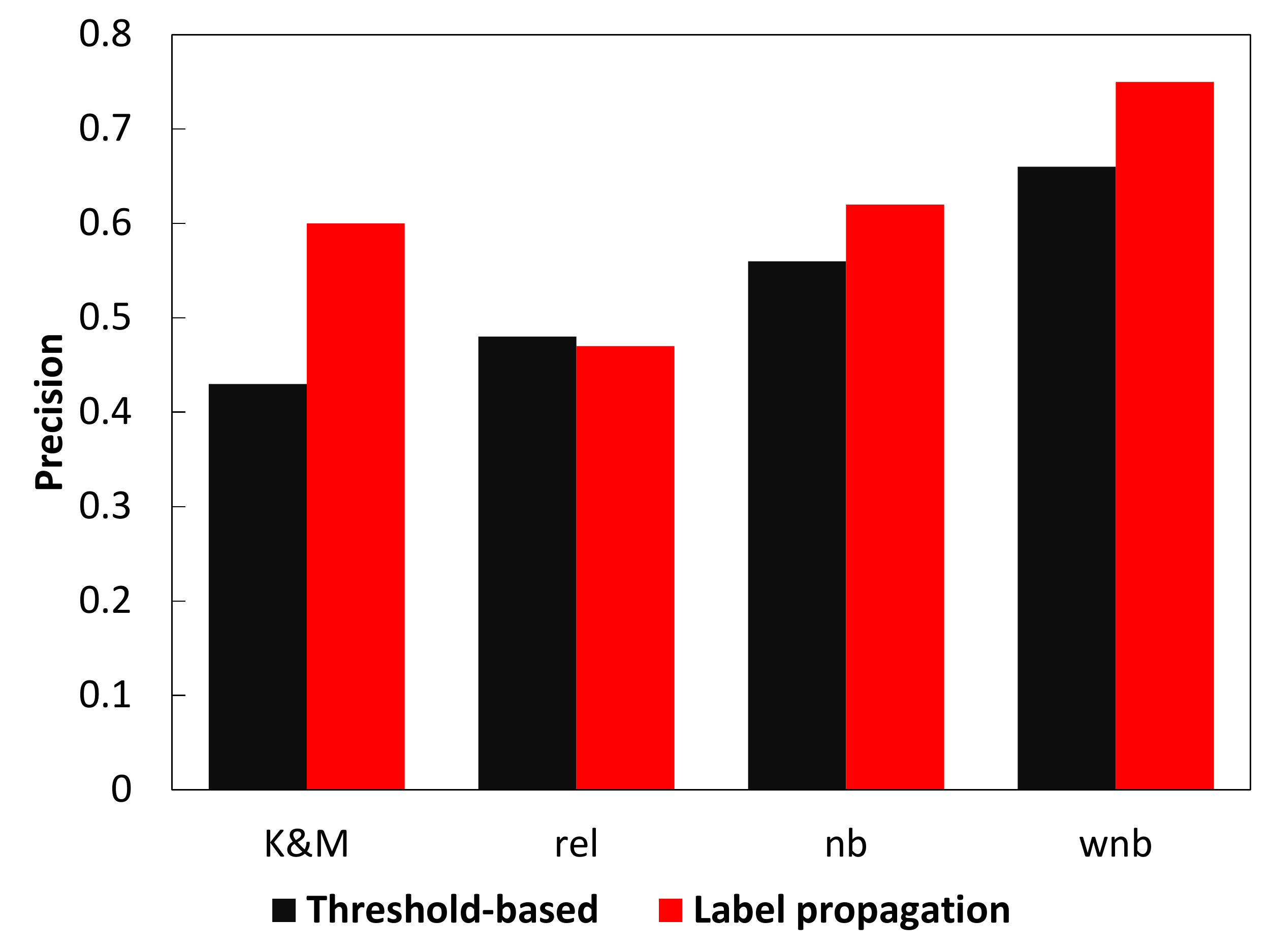}
	\caption{\small  \textmd{Comparison between threshold-based and label propagation selection approaches for the inactive class}}
	\label{fig:comp}
\end{figure}

In label propagation selection, we first select a set of users that have a high causality score as seeds, then ProSel selects users that occur with those seeds and have a score higher than a threshold iteratively. Also, the seed set in each iteration is the selected users of the previous iteration. The intuition behind this approach is to select a user if it has a score higher than a threshold or has a lower score but occurs in the cascades that high score users occur. We set the parameters of \textit{ProSel Algorithm} as follows: $\lambda = 0.1$, $\theta = 0.9$, except for relative likelihood causality, where we set $\lambda = 1$, $\theta = 9$. Table~\ref{tab:pro} shows the performance of each metric. Precision of these metrics varies from 0.47 to 0.75 and $\epsilon_{wnb}$ achieves the highest precision. Metrics $\epsilon_{rel}$ with average cascade size of 612.04 and  $\epsilon_{nb}$ with median of 230 find the more important PSM accounts. Moreover, detected PSM accounts have participated in the larger cascades compared with threshold-based selection.
 This approach also produces much lower number of false positives compared to threshold-based selection. The comparison between this approach and threshold-based selection is illustrated in Fig.~\ref{fig:comp}. From the precision perspective, label propagation method outperforms the threshold-based one. 

\begin{table}[ht]
	\centering
	\caption{\small  \textmd{Label propagation selection approach - number of selected users}}
	{
		\begin{tabular}{| p{1.1cm}| c c c c c|}
			\hline
			\textbf{Method} & \textbf{False Pos} & \textbf{True Pos} & \textbf{Precision} &\textbf{Avg CS}&\textbf{Med CS}
			\\ \hline \hline
			$All feat.$ &164,012&31,131&\textbf{0.16} & 273.21 & 176\\ \hline 
			$NoCont.$ &357,027&63,025&0.15& 262.97 &172\\ \hline	\hline
			$\epsilon_{K\&M}$ & 9,305 & 14,176 & 0.60&390.52 &179 
			\\ \hline
			$\epsilon_{rel}$ & 561 &  498 & 0.47&\textbf{612.04}&216 
			\\ \hline
			$\epsilon_{nb}$ & 1,101 & 1,768 & 0.62&403.55&\textbf{230} 
			\\ \hline 
			$\epsilon_{wnb}$ & 1,318  & 4,000 & \textbf{0.75}&355.24&183.5
			\\ \hline  \hline
	\end{tabular}}
	\label{tab:pro}
\end{table}

The number of common users selected by any pair of two metrics are also illustrated in Table~\ref{tab:pro_common}. It shows that our metrics are powerful to cover different regions of the search area and identify different sets of users. In total, 10,254 distinct active users and 16,096 inactive ones are selected. 

\begin{table}[ht]
	\centering
	\caption{\small  \textmd{Label propagation selection approach - number of common selected users}}
	{
		\begin{tabular}{| p{1cm}| c c c | c c c|}
			\hline
			\textbf{Status} & \multicolumn{3}{c|}{\textbf{Active}} &  \multicolumn{3}{c|}{\textbf{Inactive}} \\ \hline
			\textbf{Method} & \textbf{$\epsilon_{rel}$} & \textbf{$\epsilon_{nb}$}  & \textbf{$\epsilon_{wnb}$}  &  \textbf{$\epsilon_{rel}$} & \textbf{$\epsilon_{nb}$}  & \textbf{$\epsilon_{wnb}$} \\ \hline \hline
			$\epsilon_{K\&M}$ & 289 & 581 &  1,122 & 168 & 1,019 & 2,788 \\ \hline
			$\epsilon_{rel}$  &   &  15 & 6 & & 180 & 102 \\ \hline
			$\epsilon_{nb}$  &  &  & 151 & & & 833 \\ \hline  \hline
	\end{tabular}}
	\label{tab:pro_common}
\end{table}

\section{Related Work}\label{sec:rw}
To the best of our knowledge, this paper represents the first unsupervised approach on PSM detection. The majority of previous work was based on three fundamental assumptions. First, \emph{the information of the network is known}~\cite{subrahmanian2016darpa,goyal2010learning,benignitweets,abokhodair2015dissecting}. This assumption may not hold in reality. Second, they are language dependent ~\cite{subrahmanian2016darpa,morstatter2016new}. Third, the majority of botnet detection algorithms focused on bots in general. That is, they did consider the bots \emph{equally}~\cite{morstatter2016new,dickerson2014using} where in this work, we identify PSM accounts that spread viral information. Here, we review related work on identifying automatic accounts and terrorist groups. Aside from the bot detection work, our work can be compared with detection of water armies.

\noindent{\textbf{Identifying Automatic Accounts.}}  Due to the importance of the issue, DARPA conducted the Twitter bot detection challenge to identify and eliminate influential bots~\cite{subrahmanian2016darpa}. In this challenge, all teams applied supervised or semi-supervised learning approaches using the diverse sets of features. Most of the previous work extracted different sets of features (tweet syntax and semantics, temporal behavior, user profile, and network features) and conducted supervised or semi-supervised approaches~\cite{subrahmanian2016darpa,dickerson2014using,morstatter2016new}. On the other hand, here, we focus on situations where neither network information nor account related attributes and user profile information are available. Our appproach is also independent of content and language.

\noindent{\textbf{Analysis of Terrorist Groups and Detection of Water Armies.}} Terrorist groups use social media for propaganda dissemination~\cite{al2015examining}. Benigni et al.~\cite{benignitweets} conducted vertex clustering and classification to find Islamic Jihad Supporting Community on Twitter. Abdokhodair et al. \cite{abokhodair2015dissecting} studied the behaviors and characteristics of Syrian social botnet. Chen et al.~\cite{chen2013battling} found that within the context of news report comments, user-specific measurements can distinguish water army from normal users. Similarly, in~\cite{chen2013best}, Chen et al. applied user behavior and domain-specific features to detect water armies. Our work is different from them since these methods also applied features related to the accounts and network (follower/followee). However, we do not have any network information and account-related features.
\vspace*{-5mm}
\section{Conclusion}
In this paper, we conducted a data-driven study on the pathogenic social media accounts especially terrorist supporters, automatic accounts and bots. We proposed unsupervised causality based framework to detect these groups. Our approach identifies these users without using network structure, cascade path information, content and user's information. We believe our technique can be applied in the areas such as detection of water armies and fake news campaigns. Currently, we are combining this method with complementary supervised approaches~\cite{morstatter2016new} and are integrating it with LookingGlass~\cite{kim2013lookingglass} in support of the U.S. Navy's Minerva program.  LookingGlass will allow us to deploy this method for identifying PSM accounts in support of real-world information operations.\medskip

{\small
	\noindent\textbf{Acknowledgments.} Some of the authors are supported through the DoD Minerva program and AFOSR (grant  FA9550-15-1-0159).}

\bibliographystyle{IEEEtran}
{

}
\end{document}